\documentclass[12pt]{article}
\usepackage[dvips]{graphicx}
\def\thefootnote{\fnsymbol{footnote}}
\def\be{\begin{equation}}
\def\ee{\end{equation}}
\def\ba{\begin{eqnarray}}
\def\ea{\end{eqnarray}}

\newcommand{\m}{\phantom{$-$}}

\newcommand{\mc}{\multicolumn}

\begin{document}
\begin{titlepage}
\thispagestyle{empty}
\vskip0.5cm
\begin{flushright}
MS--TPI--97--13
\end{flushright}
\vskip0.8cm
\begin{center}
{\Large {\bf Block Spin Effective Action}}
\end{center}
\begin{center}
{\Large {\bf for 4d SU(2) Finite Temperature}}
\end{center}
\begin{center}
{\Large {\bf Lattice Gauge Theory}}
\end{center}
\vskip0.8cm
\begin{center}
{\large Klaus Pinn \footnote{e--mail: pinn@uni--muenster.de}
 and Stefano Vinti \footnote{e--mail: vinti@uni--muenster.de}}\\
\vskip0.8cm
{Institut f\"ur Theoretische Physik I } \\
{Universit\"at M\"unster }\\ {Wilhelm--Klemm--Str.~9 }\\
{D--48149 M\"unster, Germany }
\end{center}
\vskip2.5cm
\begin{abstract}
\par\noindent
The Svetitsky--Yaffe conjecture for finite temperature  $4d$ $SU(2)$ lattice
gauge theory  is confirmed by  observing  matching of  block spin
effective actions of the  gauge model with those of the $3d$ Ising model.
The effective action for the gauge model is defined  by blocking the
signs of the Polyakov loops with the majority rule. To compute it
numerically, we apply a variant of the IMCRG method of Gupta and Cordery.
\end{abstract}
\end{titlepage}
\setcounter{footnote}{0}
\def\thefootnote{\arabic{footnote}}
\section{Introduction}

Block spin renormalization group  \cite{blockspin}
has become an important tool
in the qualitative and quantitative understanding of
critical phenomena in classical statistical mechanics and
Euclidean quantum field theory.
As a basic ingredient, it introduces effective Hamiltonians
(actions in field theoretic language) which govern block
spin degrees of freedom. The block spins are determined
from the original degrees of freedom by
averaging them over blocks.

In principle, renormalization group (RG) solves the problems posed by
critical or nearly  critical statistical systems. Under iterated
application of the block transformation, either the correlation length
in the system becomes  small, or the flow of effective Hamiltonians
eventually reaches a fixed point which determines the universal
properties of the system.

A major drawback of the RG approach stems from the fact that effective
Hamiltonians in general contain an infinite number of couplings, in
contrast to the original Hamiltonians one starts from, which usually
contain only a small number of interaction terms. The proliferation of
couplings has a number of  consequences. It is, e.g.,  not always clear
how a certain truncation to a finite number of couplings  affects the
physical results. Furthermore, even if one relies on a certain
truncation scheme, it might be tedious to explicitly compute the
effective couplings.

This might be an explanation why the task of explicit
computation of block spin effective actions has not received
very much attention in the literature. See, however, e.g.,
Refs.~\cite{kutibock} and \cite{gottlob}.

Svetitsky and Yaffe~\cite{sy} have conjectured that  a (continuous)
deconfinement transition of a $(d+1)$--dimensional finite temperature
lattice gauge theory should be in the same universality class  as the
phase transition of a corresponding $d$--dimensional spin system. This
spin system should have the center of  the gauge group as a global
symmetry group.

The Svetitsky--Yaffe conjecture  for $SU(2)$ gauge theory at finite
temperature offers the possibility of an interesting application of the
block spin renormalization group. First, it has never been rigorously
proved that this model belongs to the Ising universality class. On the
other hand, the conjecture has been checked several times by comparison
of Monte Carlo (MC) estimates for the critical indices (which were found
in good agreement \cite{efm}),  as well as with a mean field like
analytical approach  (which gives also predictions for $SU(N)$
deconfinement temperatures \cite{bcdp}). However, so far there have been
no numerical attempts to explicitly compute the effective action for the
Polyakov loops and compare it with that of the Ising model.

With this article, we intend to fill this gap.
We will demonstrate that actions for the degrees of freedom relevant for
the deconfinement transition can  well be computed by Monte Carlo.
Comparing them with the corresponding actions for the Ising model we are
able to confirm the validity of the Svetitsky--Yaffe conjecture in a
very fundamental way.

The article is organized as follows:
In Section~2 we introduce the notations for finite temperature lattice gauge
theory and recall the Svetitsky--Yaffe conjecture.
In Section~3 we introduce the block spin renormalization group.
Section~4 explains the idea of flow matching, and
the application of Improved Monte Carlo Renormalization Group (IMCRG)
\cite{gc}
to $SU(2)$ lattice gauge theory is described in some detail.
In Section~5 we discuss some details of our Monte Carlo
methods and present the results.
Conclusions follow.
\section{Finite Temperature Lattice Gauge Theory}

Let us briefly review the formulation of finite temperature gauge
theory on a lattice (see for instance Ref.~\cite{bs}).

Consider an $SU(N)$ gauge system on a $(d+1)$--dimensional hypercubic lattice
of size $L^d\cdot N_T$, where $L$ and $N_T$ are the spatial and temporal
extensions, respectively, in units of the lattice spacing $a$.

A Euclidean quantum field theory at finite temperature is obtained if one
compactifies the (imaginary) temporal direction, keeping infinite
the spatial directions.
In a finite lattice formulation one therefore assumes $L\gg N_T$.
The compactification length is proportional to the inverse physical
temperature $T$
\be
N_T\cdot a = \frac{1}{T} ~~ .
\ee
Denote with $U_\mu(n)$ the $SU(N)$ group element belonging to the
link with origin in the site $n\equiv ({\bf x}, t)$ and pointing in the
$\mu$--direction.
The usual Wilson action  reads
\ba
S_g(U) &=& \beta \sum_{P} \left(N - {\rm Re} ~{\rm Tr} U_P\right) \, ,\\
\beta  &=& \frac{2N}{g^2} ~a^{d-1} \, ,
\ea
where $U_P$ is the product of the group elements around the plaquette $P$.
The partition function is given by
\be
Z = \int\prod_{n,\mu} d U_\mu(n) \exp\left[-S_g(U)\right] \, .
\ee
Because of the periodicity in the temporal direction, the system is also
invariant under a global $Z_N$ symmetry, i.e.\ the center of the
gauge group: its spontaneous symmetry breaking at a finite temperature $T_c$
is the signal of the deconfinement transition.

The Polyakov loop is an order parameter for the finite temperature
deconfinement transition.
It is the trace of the ordered product of all
timelike links with the same space coordinate, wrapping in the time direction
\be
{\cal L}({\bf x})=\mathrm{Tr}\prod_{t=1}^{N_T}U_0({\bf x},t) ~~.
\ee
It  is a non--trivial observable from
a topological point of view:
its vacuum expectation value is not invariant under $Z_N$ transformations.
It is zero in the confining phase, while it acquires a finite value in the
deconfined phase.

According to the 15 years old Svetitsky--Yaffe conjecture~\cite{sy},
integrating out the space--like degrees of freedom one should obtain
an effective action for the Polyakov loops which is short ranged and
has the center of $SU(N)$ as a global symmetry group.

Thus, given a $d$--dimensional classical spin system with the  same
symmetry properties, undergoing a continuous phase transition, the
$(d+1)$--dimensional quantum gauge model  is expected to be in its
universality class if the deconfinement transition is a continuous one
and the  effective Hamiltonian has good locality properties.

This applies in particular to the 4--dimensional $SU(2)$ gauge model
which should belong to the $3d$ Ising universality class.
\section{Block Spin Renormalization Group}

To define the block spin  transformation, consider a magnetic system
consisting of spins $\sigma$ on the sites of a $d$--dimensional lattice,
defined by a  Hamiltonian $H$ and a set of couplings $\{K\}$,
\be
H(\sigma) = - \sum_\alpha K_\alpha S_\alpha (\sigma)~~.
\ee
The partition function reads
\be
Z = \sum_{\{\sigma\}} \exp{\textstyle\left[-H(\sigma)\right]}~~.
\ee
The ``operators'' $S_\alpha(\sigma)$ are in general all possible
products of spins compatible with the symmetry of the Hamiltonian.
Explicit examples will be given below.

A block spin transformation maps the fine $L^d$ lattice onto the  block
lattice of size $L'^d$, where $L=L_B \, L'$.   This is achieved by
averaging the original spins over cubical blocks of side length
$L_B$  according to a certain rule.

The Hamiltonian $H'$ of the block spins $\{\mu\}$
assigned to the sites of the block lattice
is defined by
\be
\exp{\textstyle\left[-H'(\mu)\right]} = \sum_{\{\sigma\}} P(\mu,\sigma) \,
\exp{\textstyle\left[-H(\sigma)\right]} \, ,
\label{effec}
\ee
where $P$ encodes the mapping from the
fine to the coarse lattice. It obeys
\be
P(\mu,\sigma) \geq 0 \; \mbox{\ and \ } \sum_{\{\mu\}} P(\mu,\sigma) = 1~.
\ee
The latter property ensures that the partition function remains
unchanged,
\be
Z = \sum_{\{\mu\}} \exp{\textstyle\left[-H'(\mu)\right]} \, .
\label{Zblock}
\ee
In this work we use the majority rule prescription (i.e. the $\mu$ spins
take values plus or minus one)
\be\label{pdef}
P(\mu,\sigma)
= \prod_{{\bf x'}}^{\rm (blocks)} \frac{1}{2} \left[ 1 + \mu_{{\bf x'}}~
\mbox{sign} \sum_{{\bf x}\in {\bf x'}} \sigma_{\bf x} \right] ~~~.
\label{majo}
\ee
The sign function
sign$(x)$ in Eq.~(\ref{pdef}) is defined such that it vanishes for $x=0$.
This ensures that in case of a zero sum of spins
inside a block  a positive (negative) $\mu_{\bf x'}$ is selected with
probability one half.

The block Hamiltonian $H'$ can be expressed in terms of
operators $S_\alpha'(\mu)$, defined on the block lattice,
\be
H'(\mu) = - \sum_\alpha K'_\alpha S_\alpha'(\mu)  \, .
\label{newH}
\ee

In the case of the 4--dimensional
$SU(2)$ gauge model, the $3d$ effective action
for the signs of the Polyakov loops
shares (by definition) the $Z_2$ symmetry with
the $3d$ Ising model.
To define this action we assign to each Polyakov loop its sign
\be
\sigma_{{\bf x}}(U)={\rm sign}~{\cal L}({{\bf x}}) \, .
\label{mapping}
\ee
Then we block the $\sigma$--spins with the majority rule to obtain
Ising type block $\mu$--spins.

It follows that, similarly to Eq.~(\ref{effec}), the effective Hamiltonian
$\mathcal{H'}$ for the finite temperature gauge system is given by
\be
\exp[-\mathcal{H'}(\mu)] = \int DU~P(\mu,U)~\exp[-S_g (U)] \, ,
\label{blockg}
\ee
with
\be
P(\mu,U)
= \prod_{{\bf x'}}^{\rm (blocks)} \frac{1}{2} \left[ 1 + \mu_{{\bf x'}}~
\mbox{sign} \sum_{{\bf x}\in {\bf x'}} \sigma_{\bf x}(U) \right]~~.
\label{majog}
\ee
Other procedures of blocking, like
first averaging the Polyakov loops inside the blocks and then take as
Ising type spin its sign, could also be employed.

We close this section by defining a renormalization group flow.  A
natural way to do it would be to fix a block length, e.g., $L_B=2$,  and
then iterate the transformation (\ref{majog}). We do {\em not}  stick to
this definition here. Instead we define the flow  by just increasing the
block size $L_B$.  This allows us to compute Hamiltonians not only for
scales  $2^n$, but also for arbitrary scales $L_B$, with
$L_B$ integer.
\section{Monte Carlo Renormalization Group for Polyakov Loops}

In this section we  first recall the RG matching idea. Then we show how
to apply   the Improved Monte Carlo
Renormalization Group (IMCRG)
method by Gupta and Cordery~\cite{gc} to $4d$
$SU(2)$ gauge theory at finite temperature.
\subsection{Matching of RG Trajectories}

In the infinite--dimensional space of couplings $\{K\}$,  a
renormalization group  transformation $R$ can be  looked at as a mapping
of the original bare Hamiltonian $H$ onto a new Hamiltonian $H'=R(H)$,
defined by the couplings $\{K'\}$.

The RG flow obtained under iterated  RG transformations  will eventually
end in a fixed point $\{K^*\}$, defined  through $H^*=R(H^*)$. The
critical surface is identified by all flows connected to the fixed point
in this way.

The RG matching method is based on the  assumption that different
physical systems, belonging to the same universality class, will follow
RG flows which originate from different ``bare couplings'' on the
critical surface and eventually match in a neighbourhood of the common
fixed point. Of course, a matching close to a non--trivial fixed point
will  only take place if both models under consideration are at
criticality. A matching thus confirms both universality and  allows to
check for criticality.  This matching method has been successfully
applied in the context of spin  models, see e.g.~\cite{hp}.
\begin{figure}
\begin{center}
\includegraphics[width=9cm]{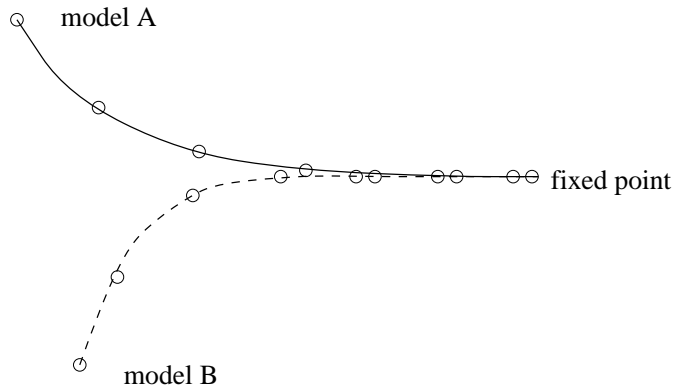}
\parbox[t]{.85\textwidth}
 {
 \caption[matchi]
 {\label{matchi}
\small
 Matching of the RG trajectories of two critical models A and B in
 the neighbourhood of an RG fixed point.
 }
 }
\end{center}
\end{figure}
The feasibility of matching different critical flows by means of  MC
methods mainly relies on the assumption that the different trajectories
come close to each other, i.e.\ approximately match, before the fixed
point is actually reached, c.f.\  Figure~\ref{matchi}.  As we shall see,
this condition is met for our models.
\subsection{Effective Couplings from IMCRG}

The Improved Monte Carlo Renormalization Group  method~\cite{gc} allows
to  compute effective actions for Ising type block spins.\footnote{A
generalization to systems with continuous block variables is not
straightforward.}

The main idea is to avoid simulations of the original partition function.
Instead, consider a
{\em modified} system defined through
\ba
Z_c &=&\sum_{\{\mu\}} ~
\exp\left[ - H'(\mu) \; + \; \bar H(\mu)\right]~ \nonumber  \\
 &=& \sum_{\{\mu\}} \sum_{\{\sigma\}} ~ P(\mu,\sigma) ~
\exp\left[ - H(\sigma) \; + \; \bar H(\mu)\right] \, ,
\label{compensated}
\ea
where
\be
\bar H(\mu)= - \sum_{\alpha} \bar K'_\alpha S_{\alpha}'(\mu)
\label{guessh}
\ee
is a {\em guess} for $H'(\mu)$.

This system can be simulated once $H$ and $\bar H$ are given.
Note the {\em plus} sign in front of $\bar H$ in
Eq.~(\ref{compensated}).

The system with partition function $Z_c$ is
{\em non--critical}, even in the case of a critical Hamiltonian
$H(\sigma)$.
Consider the expectation values
\ba
<S_\alpha'>_c ~&=& ~ \frac{1}{Z_c} \sum_{\{\mu\}}
\, S_\alpha' ~ e^{\textstyle ~\left[-H'(\mu)+\bar H(\mu)\right] } \\
&=& ~  \frac{1}{Z_c} \sum_{\{\sigma\}}\sum_{\{\mu\}}
\, S_\alpha' ~P(\mu,\sigma) ~e^{\textstyle ~\left[-H(\sigma)+\bar H(\mu)
\right] }~~.
\ea
If the guess is exact, i.e.,
\be
\bar H(\mu) = H'(\mu) \, ,
\label{exguess}
\ee
the block spins $\mu_{\bf x'}$ completely decouple and fluctuate independently.
In other words, the system is non--critical and the correlations in
$Z_c$ are bounded by the block size $L_B$.
The correlations functions are then known exactly,
\ba
<S_\alpha'>_o &=& 0 \, , \nonumber  \\
<S_\alpha' S_\beta'>_o &=& n_\alpha~\delta_{\alpha\beta} \, ,
\ea
where $n_\alpha$ are trivial
multiplicity factors.

Let us assume that
$\bar H(\mu)$ is close to $H'(\mu)$. Then
a first order expansion gives
\be
<S_\alpha'>_c = n_\alpha~ (K'_\alpha-\bar K'_\alpha) ~ + ~
O \left( (K'_\alpha-\bar K'_\alpha)^2 \right) \, .
\ee
Solving this equation for $K'_\alpha$ allows to improve
the guess $\bar K'_\alpha$. Usually a few iterations
\be
\bar K'_\alpha  \rightarrow \bar K'_\alpha + n_\alpha^{-1}<S'_\alpha>_c \, ,
\label{recursion}
\ee
where the expectation values are determined by simulation
of the system~(\ref{compensated}),
are sufficient to determine $H'$ to a good precision.

To apply the IMCRG procedure to the $SU(2)$ gauge system, one
has to simulate the partition function
\ba
Z_c &=&~ \sum_{\{\mu\}} ~
\exp\left[ - \mathcal{H'}(\mu) \; + \; \bar H(\mu)\right] \nonumber \\
 &=&~ \sum_{\{\mu\}} \int DU~ P(\mu,U)~
\exp\left[ - S_g (U)\; + \; \bar H(\mu)\right] \, .
\label{compensatedG}
\ea
Remember that the $\mu$--variables are defined as the majority
rule block spins of the signs of the Polyakov loops.
It is straightforward to design a MC procedure for the updating
of this system. Details will be given in the next section.
\section{Monte Carlo Simulations}

We applied the IMCRG method to three different systems: the $3d$ standard
Ising model (with nearest--neighbour coupling),  the $3d$ ``$I_3$'' model,
which includes also a third, cube--diagonal  neighbour
coupling~\cite{blh}, and the $4d$ $SU(2)$ pure gauge model. We
simulated the system defined by  Eq.~(\ref{compensated}) for
the spin models and  by Eq.~(\ref{compensatedG}) for the $SU(2)$ gauge
model.

For the updating of the Ising model we used a Metropolis algorithm:
A single spin $\sigma_{\bf x}$ is proposed to be flipped.
It is checked whether
this update leads to a flip of the block spin $\mu_{\bf x'}$, with
${\bf x} \in {\bf x'}$.
The total change of energy $\Delta H(\sigma) - \Delta \bar H(\mu)$
is then computed and used in the
usual Metropolis acceptance/rejectance step.

In case of the $SU(2)$ model, only the temporal links couple to the
compensating block Hamiltonian.  The space--like links are updated using
the incomplete  Kennedy--Pendleton heat bath sweep \cite{kennedy}
supplied with a
number of overrelaxation sweeps. For temporal links one  employs again a
Metropolis procedure: A proposed change of a link matrix leads to a
change of the Polyakov loop $\cal L({\bf x})$
of which it is member. If the sign
$\sigma_{\bf x}$ changes, the block spin $\mu_{\bf x'}$ in turn might
flip and give rise to a change of  $\bar H(\mu)$. The relevant energy
change for the Metropolis step is $\Delta S_g(U) - \Delta \bar H(\mu)$.

\begin{figure}
\begin{center}
\includegraphics[width=8cm]{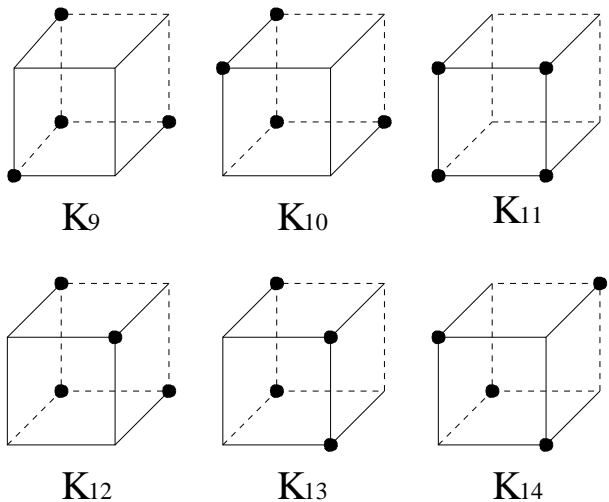}
\parbox[t]{.85\textwidth}
 {
 \caption[quart]
 {\label{quart}
\small
Graphical definition of 4--spin couplings included in the
effective Hamiltonian.
 }
 }
\end{center}
\end{figure}

In practical calculations one has to truncate the interactions in $H'$
and $\bar H$. We chose to include in the ansatz eight 2--point couplings
and six 4--point  couplings.  The 2--point couplings can be labelled by
specifying the relative position of the interacting spins (up to obvious
symmetries):  Our couplings $K_1 \dots K_8$ then correspond to 001, 011,
111, 002, 012, 112, 022, 122.  The 4--point couplings $K_9 \dots K_{14}$
are  defined in an obvious way through  Figure~\ref{quart}.  The
corresponding interaction terms in the effective Hamiltonian are denoted
by $S_\alpha'$, $\alpha=1 \dots 14$.
\begin{figure}
\begin{center}
\includegraphics[width=6cm]{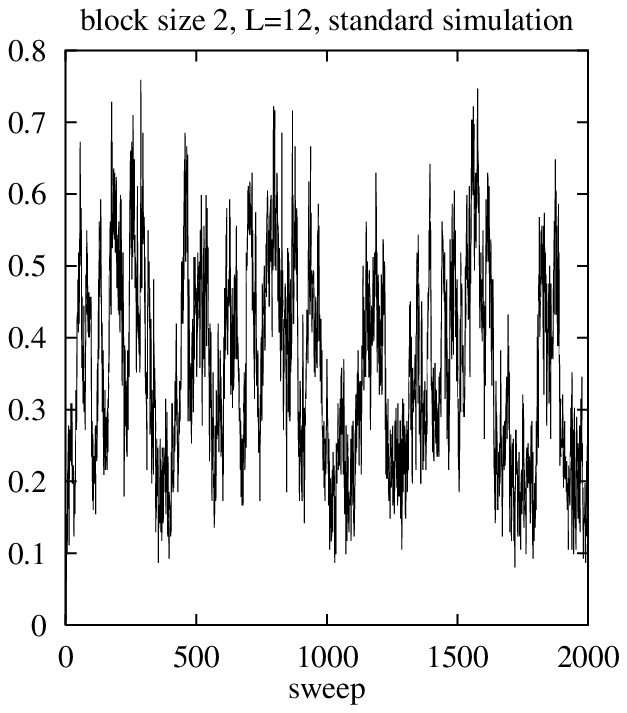}\includegraphics[width=6cm]{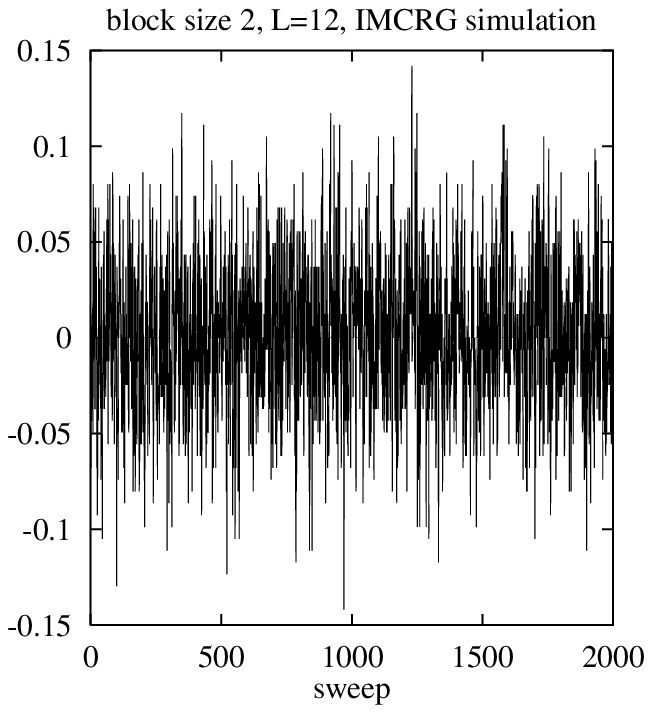}
\vspace{1cm}
\includegraphics[width=6cm]{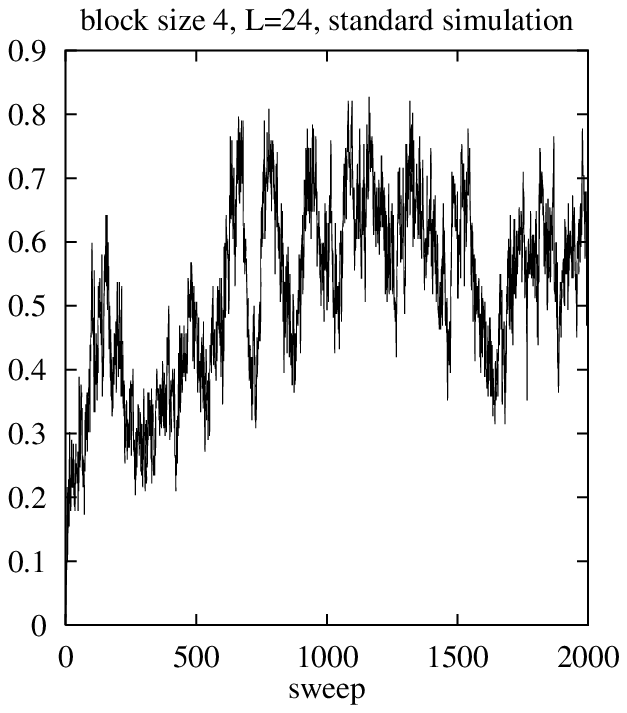}\includegraphics[width=6cm]{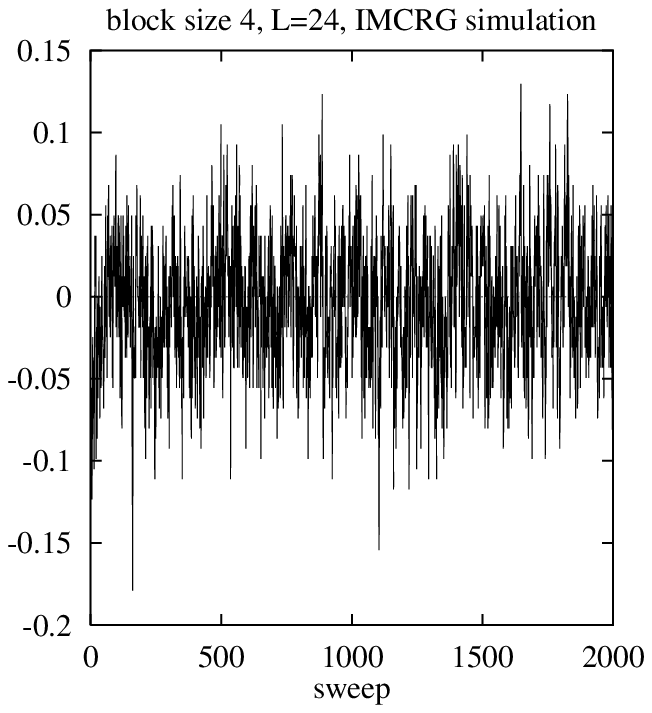}
\parbox[t]{.85\textwidth}
 {
 \caption[csd]
 {\label{csdfig}
\small
 Comparing scatter plots of the nearest neighbour
 block correlation function
 observable in $SU(2)$ simulations with $N_T=2$.
 The standard simulations (no compensation on block level) are
 shown on the left, the IMCRG simulations on the right.
 }
 }
\end{center}
\end{figure}
\subsection{Reduction of Critical Slowing Down}

A merit of the IMCRG method is that
block spin observables are (nearly) decorrelated and
the critical slowing down problem is less severe than
in standard simulations.

In Figure~\ref{csdfig}
we show scatterplots (MC time history) of measurements of the
nearest neighbour block spin 2--point function. The comparison is
between a simulation of the pure gauge system (without IMCRG
compensation on the block level)  and the system defined through
Eq.~(\ref{compensatedG}). The plot clearly shows that the IMCRG type
simulation suffers much less from critical slowing down.

Analogous observations were made in case of the Ising model
simulations with the compensating block Hamiltonian switched on.
It is the reduction of critical slowing down
obtained from the compensation even on moderate lattice sizes which
enabled us to obtain reasonable results with moderate CPU
expense.
\subsection{Matching of the Two Ising Models}

We started by comparing the RG flow of the two
Ising models.
In Table~\ref{tab1} we  summarize some parameters of the MC simulations.
We made simulations on lattices consisting of $8^3$ blocks of size $L_B$
at the infinite volume critical couplings
$\beta=0.2216544$~\cite{isingbeta} for the ordinary Ising and
$(\beta_1,\beta_3)=(0.128003,0.051201)$~\cite{blh} for the $I_3$ model.
\begin{table}
\small
\begin{center}
\begin{tabular}{lcl}
\hline
\hline
&   &  \\[-3mm]
 model & couplings & \mc{1}{c}{$L_B$} \\[0.5mm]
\hline
&  &  \\[-3mm]
 standard Ising & \phantom{(} 0.2216544 & 3,5,7,9,13    \\
 $I_3$ Ising    & (0.128003, 0.051201) &  3,5,7,9,13,17 \\[1mm]
 \hline
 \hline
 \end{tabular}
\parbox[t]{.85\textwidth}
 {
 \caption[Table 1]
 {\label{tab1}
\small
Block sizes $L_B$ and couplings used for the IMCRG simulations of the
standard and $I_3$ Ising models.
For all runs we used $L'=8$.
}
}
\end{center}
\end{table}
At each RG step (fixed $L_B$ value) we made
usually two, three or four IMCRG iterations in order to
have the guesses of the effective couplings converge
to reasonable precision.
The number of sweeps in each run ranged from
$\sim 10^5$ for the  largest lattice sizes to a
$\sim 6\cdot 10^6$ for the smallest ones.

As the final  error of the estimate for an effective coupling, we took
the maximum of the statistical error and the last change of
guess in the IMCRG procedure.
\begin{figure}
\begin{center}
\includegraphics[width=7cm]{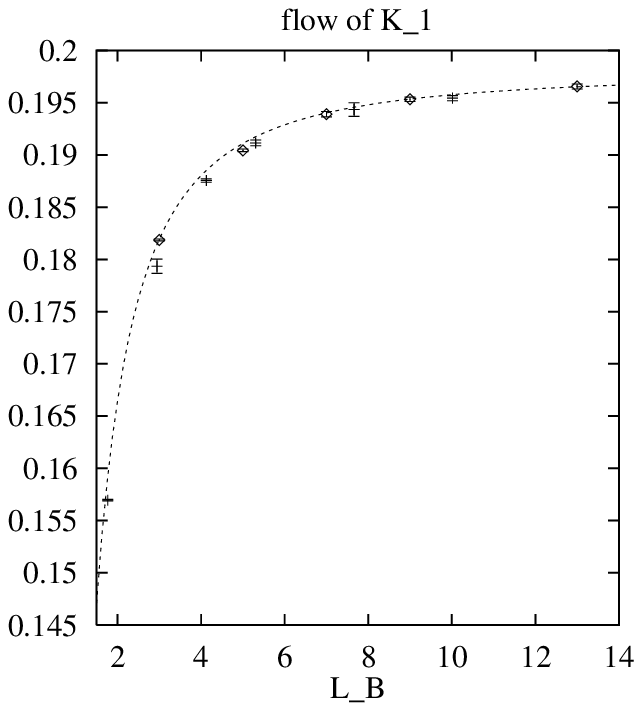}
\vspace{1cm}
\includegraphics[width=7cm]{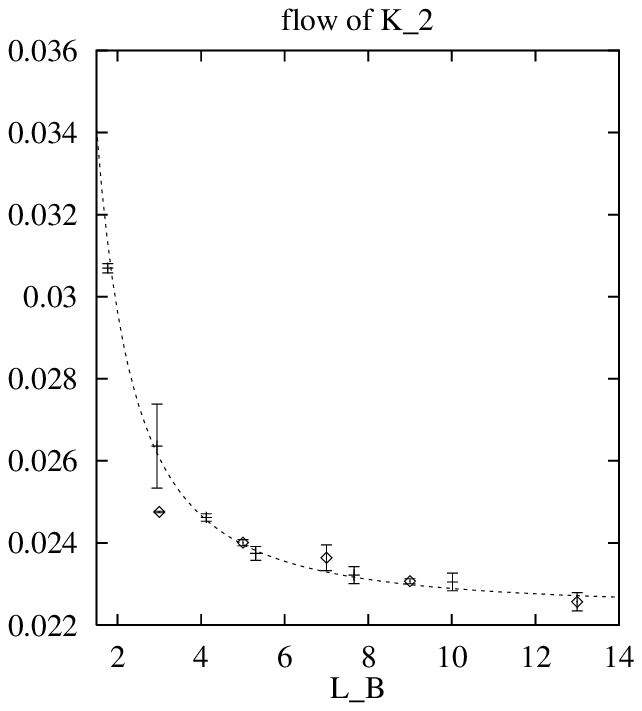}
\parbox[t]{.85\textwidth}
 {
 \caption[isiflow]
 {\label{isiflow}
\small
 Flows of nearest and second nearest neighbour couplings in the standard
 (diamonds) and the $I_3$ (bars) Ising model with increasing block
 size $L_B$. In order to obtain matching,
 the block sizes of the $I_3$ model were rescaled by a factor
 $\lambda=0.59$. The dotted lines are fits of the flows with a power law.
 }
 }
\end{center}
\end{figure}
In Figure~\ref{isiflow} we show the results for the flow of the  two
leading couplings $K'_1$, and $K'_2$, with increasing block size $L_B$.
To achieve matching, the block sizes $L_B$ of the $I_3$ model have been
rescaled by a factor of $\lambda=0.59$.  A rescaling is always needed to
obtain matching. The reason for this is that  the two models
have a different distance to ``travel'' before they meet on a common
trajectory, c.f.\ Figure~\ref{matchi}.

The figure shows that the two flows collapse nicely on a  single
trajectory, indicating that they are approaching a common fixed point.
This happens also with the other 12 couplings not shown in the plot. It
turns out that the approach to the fixed point can be well fitted by a
power law,
\be\label{fit}
 K_\alpha(L_B) = K_\alpha^{*} + a_\alpha \cdot L_B^{-\rho}\,.
\ee
If the flows of two models obey such a law, with the same
fixed point  $K_\alpha^{*}$ and exponent $\rho$, but
different ``amplitudes'' $a$ and $a'$,
the rescale
factor $\lambda$ to obtain matching is
\be\label{amplitu}
\lambda = ~\left(\frac{a'}{a}\right)^{\frac{1}{\rho}} \, .
\ee
\begin{table}
\small
\begin{center}
\begin{tabular}{lclllc}
\hline
\hline
&  &  &  &  &  \\[-3mm]
  \mc{1}{l}{\phantom{1}~$\alpha$, model} &
  \mc{1}{c}{$L_B\ge$} &
  \mc{1}{c}{$K_\alpha^{*}$} &
  \mc{1}{c}{$a_\alpha$} &
  \mc{1}{c}{$\rho$} &
  \mc{1}{c}{$\chi^2/$dof} \\[1mm]
\hline
&  &  &  &  &  \\[-3mm]
001, Ising   & 3 & \m 0.1990(5) & $-$0.077(4) & 1.37(7)   & 0.99 \\
001, Ising   & 5 & \m 0.1977(7) & $-$0.15(7)  & 1.9(3)    & 0.01 \\
001, $I_3$   & 3 & \m 0.1981(4) & $-$0.240(8) & 1.60(4)   & 0.50 \\
001, $I_3$   & 5 & \m 0.1975(5) & $-$0.33(9)  & 1.80(14)  & 0.09 \\
001, combined& 7 & \m 0.1979(6) & $-$0.10(3)  & 1.67(20)  & 0.17 \\
             & 7 &              & $-$0.27(9)  &           &      \\[0.5mm]
\hline
&  &  &  &  &  \\[-3mm]
011, Ising   & 5 & \m 0.0224(1) & \m 0.023(2) & 1.67 fix & 0.94 \\
011, $I_3$   & 5 & \m 0.0225(2) & \m 0.055(6) & 1.67 fix & 0.27 \\[0.5mm]
\hline
&  &  &  &  &  \\[-3mm]
111, Ising   & 5 & \m 0.0013(1) & \m 0.015(2) & 1.67 fix  & 0.15 \\
111, $I_3$   & 5 & \m 0.0013(1) & \m 0.029(2) & 1.67 fix  & 1.12 \\[0.5mm]
\hline
&  &  &  &  &  \\[-3mm]
002, Ising   & 5 & $-$0.0202(1) & \m 0.047(2)  & 1.67 fix  & 1.53 \\
002, $I_3$   & 5 & $-$0.0201(3) & \m 0.059(8)  & 1.67 fix  & 0.38 \\[0.5mm]
\hline
&  &  &  &  &  \\[-3mm]
{\phantom{10}9}, Ising & 5 & \m 0.00210(4) & $-$0.006(1) & 1.67 fix  & 0.10 \\
{\phantom{10}9}, $I_3$ & 5 & \m 0.00215(10)& $-$0.005(3) & 1.67 fix  & 0.17
\\[1mm]
\hline
\hline
\end{tabular}
\parbox[t]{.85\textwidth}
 {
 \caption[Table 2]
 {\label{tab2}
\small
Fit results for the Ising model flows for a number of 2--point couplings
and for the largest 4--point coupling $K_9$. The fits were done with
Eq.~(\ref{fit}). The second column gives the minimum block sizes
that were used in the fit.
$K^{*}_\alpha$ are the estimates for the fixed
point values.
A ``fix'' after a parameter means that the value was kept fixed
during the fitting procedure.
}
}
\end{center}
\end{table}
We always used as a reference
trajectory  the flow of the standard Ising model
and rescaled the block sizes of the other models by
an appropriate factor.

The results of our various power law fits of the Ising model
are summarized in Table~\ref{tab2}. We fitted the models
separately, checking also the effect of discarding the
effective couplings for the smallest block sizes.
That the fixed point value and the
exponent of the two models coincide is confirmed by
a common fit of the two Ising flows, where in the
fit function only the amplitudes $a_\alpha$ were allowed
to depend on the model. This yields the results quoted
in the last two lines of the first block of the table.
We then fixed $\rho=1.67$ and fitted the flows of the non--leading
couplings with two parameters (fixed point value and
amplitude of power law correction). We found a very nice agreement
of the resulting fixed point values for all couplings.

The value of the  exponent $\rho$ turns out to be too big
to be identified with the first correction to scaling exponent
$\omega\approx 0.8$. We expected that $\omega$ should be the  leading
exponent. A possible explanation of the present observation  is the
following: The amplitude of a power term with $\omega$ as  exponent is
too small to be detected within our precision.  The exponent $\rho$ with
its relatively large amplitudes is due to the presence  of a redundant
operator of the particular blocking scheme we used.
\subsection{Matching of the 4d SU(2) with the Ising Models}

We then turned to the $4d$ $SU(2)$ gauge model at finite temperature.
Informations on MC simulations made in this case are given in
Table~\ref{tab3}.
\begin{table}
\small
\begin{center}
\begin{tabular}{cllc}
\hline
\hline
 &  &   &  \\[-3mm]
 $N_T$ & \mc{1}{c}{$\beta$} & \mc{1}{c}{$L_B$} & $L^3\cdot N_T$\\[0.5mm]
\hline
 &  &   &  \\[-3mm]
  1 &0.8730 & 2,3,4,5,6,7&  $42^3\cdot 1$\\
  2 &1.871  & \phantom{2,}3,4,5,6  &  $36^3\cdot 2$\\
  2 &1.874  & \phantom{2,}3,4,5,6  &  $36^3\cdot 2$\\
  2 &1.877  & 2,3,4,5,6  &  $36^3\cdot 2$\\
  2 &1.880  & 2,3,4,5,6  &  $36^3\cdot 2$\\[1mm]
\hline
\hline
\end{tabular}
\parbox[t]{.85\textwidth}
 {
 \caption[Table 3]
 {\label{tab3}
\small
Lattice sizes and values of $\beta$ used for the
{\rm $4d$ $SU(2)$} model at finite temperature.
In the last column the maximum lattice size is shown.
For all runs we used $L'=6$.
}
}
\end{center}
\end{table}
We made MC simulations for the $N_T=1$ and $N_T=2$ cases on lattices
consisting of $6^3$ blocks of size $L_B$.

As the critical deconfinement transition value we used the gauge couplings
$\beta_c=0.8730(2)$~\cite{bems} for $N_T=1$.
For $N_T=2$ we studied a neighbourhood of the critical value
$\beta_c=1.880(3)$~\cite{fhk} (see Table~\ref{tab3}).

For the $SU(2)$ model statistic has necessarily been reduced compared
to the Ising models.
For $N_T=2$ measurements ranged from $\sim 10^4$ for the largest sizes up to
$\sim 5\cdot10^5$ for the smaller ones.

\begin{figure}
\begin{center}
\includegraphics[width=10cm]{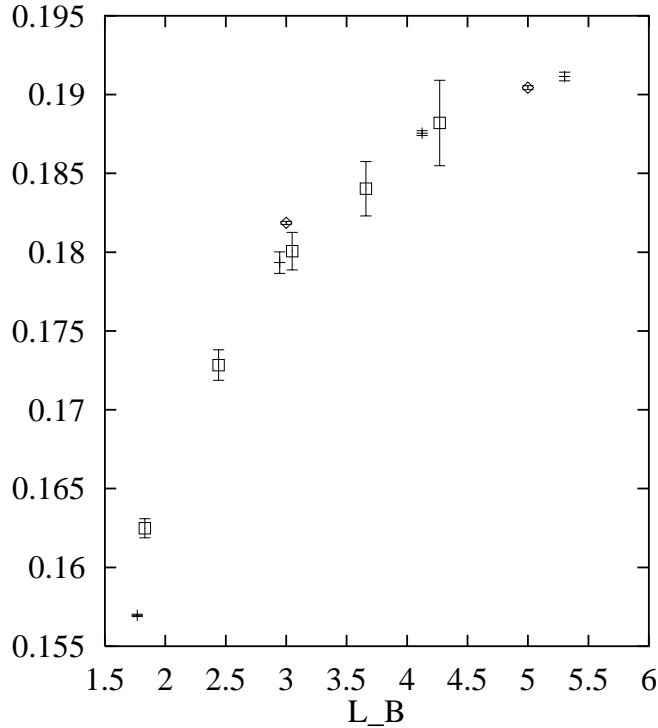}
\parbox[t]{.85\textwidth}
 {
 \caption[nt1flow]
 {\label{nt1flow}
\small
 Flow of the nearest neighbour coupling in the effective action for
 $N_T=1$ lattice gauge theory with gauge coupling $\beta = 0.8730$
 (squares). Also shown is the standard Ising (diamonds)
 and  and $I_3$ model (bars).
 The rescaling factor of the gauge block size with respect to
 the standard Ising scale is $\lambda=0.61$.
 }
 }
\end{center}
\end{figure}

The $K'_1$ leading coupling result for $N_T=1$ is shown in
Figure~\ref{nt1flow}.
Also shown is the Ising flow.
Notice that here and in the figures which follow the Ising flows are
the same as in Figure~\ref{isiflow}, which also means that the fit lines
plotted are those obtained from the Ising data.
\begin{figure}
\begin{center}
\includegraphics[width=6cm]{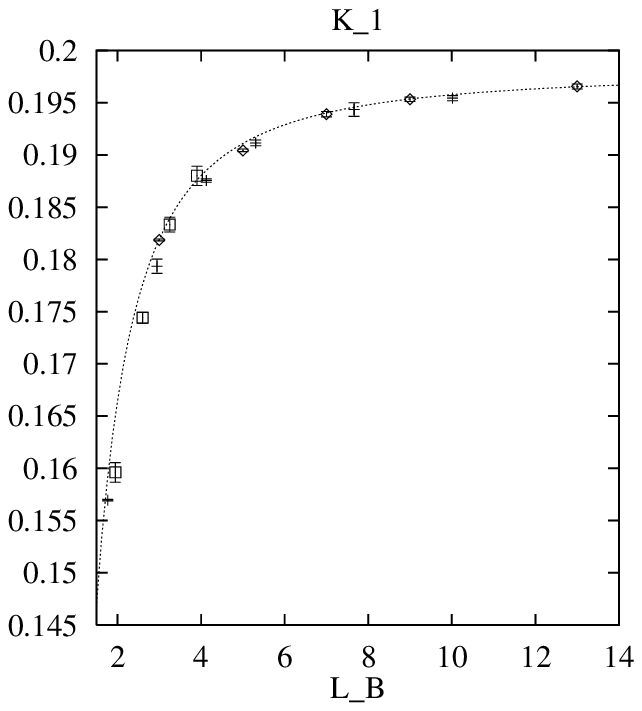}\includegraphics[width=6cm]{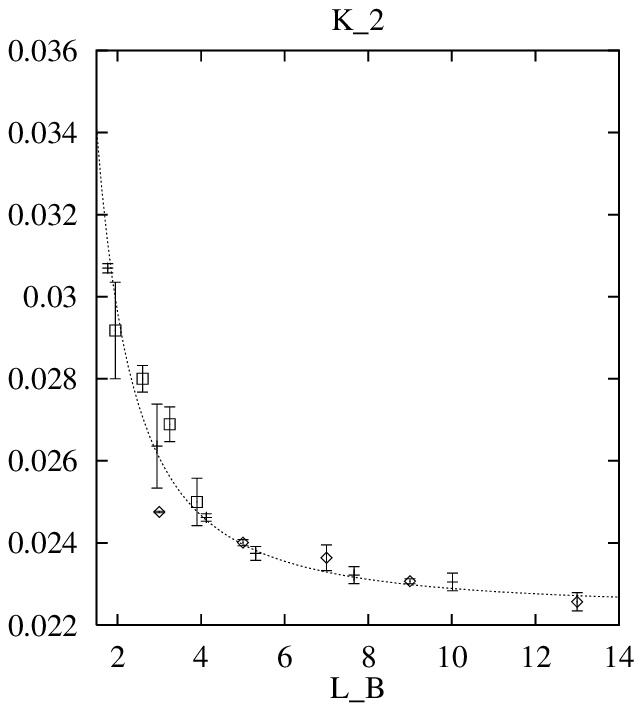}
\vspace{1cm}
\includegraphics[width=6cm]{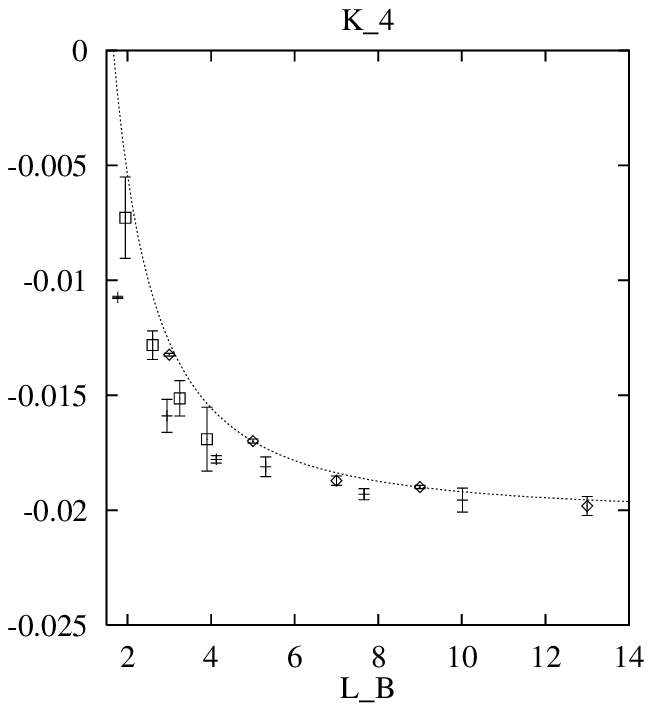}\includegraphics[width=6cm]{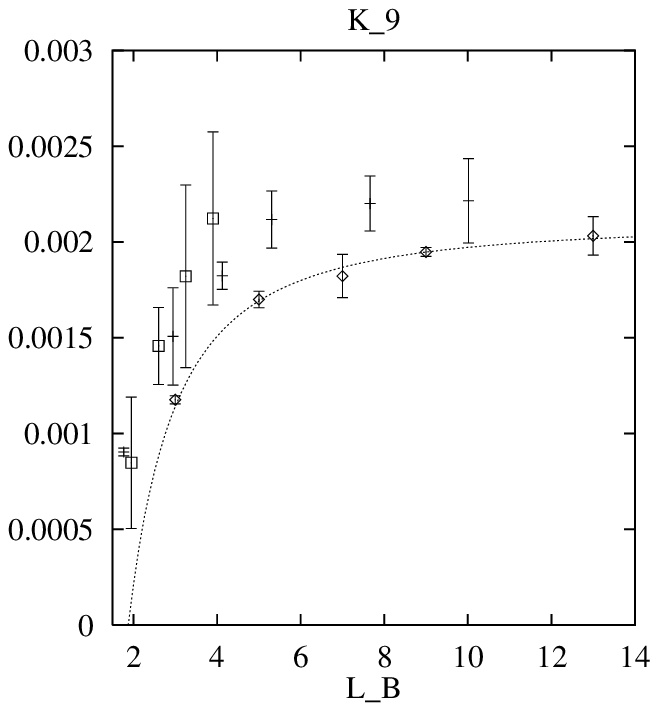}
\parbox[t]{.85\textwidth}
 {
 \caption[nt2flow]
 {\label{nt2flow}
\small
 Flows of four different effective couplings for $N_T=2$ $SU(2)$ lattice
 gauge theory at $\beta=1.877$, matching with the two Ising models
 (bars, diamonds, and fit lines).
 The block sizes of the gauge model are rescaled by a factor
 $\lambda=0.65$ with respect to the standard Ising scale.
 }
 }
\end{center}
\end{figure}
Fits were made again according to Eq.~(\ref{fit}).
Omitting the $L_B=2$ value, they gave results consistent with the fits of
the Ising data, both for the exponent and the asymptotic values,
however with bigger errors due the lower statistics.
The  rescaling of the gauge block sizes with respect to the standard Ising
scale used in this case is $\lambda=0.61$.

The flows of the effective couplings for $N_T=2$ at
$\beta=1.877$ are given in Figure~\ref{nt2flow}.
One can see a clear matching
with the two Ising models (bars and diamonds) and the fit lines for the
first two couplings.
The $L_B$'s of the $SU(2)$ gauge model were rescaled in this case by a factor
of $\lambda=0.65$.

\begin{figure}
\begin{center}
\includegraphics[width=10cm]{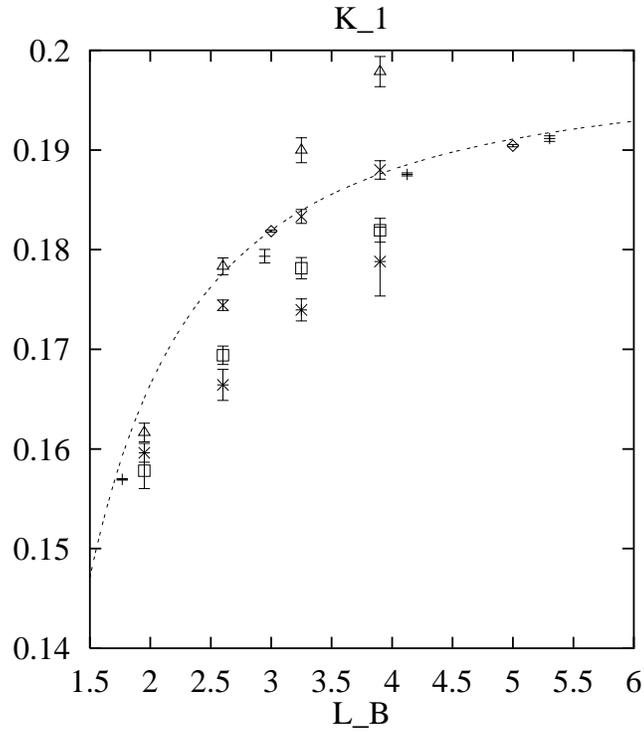}
\parbox[t]{.85\textwidth}
 {
 \caption[allflow]
 {\label{allflow}
\small
Flows of the NN coupling in the effective action for
$N_T=2$ $4d$ $SU(2)$ at four different gauge couplings $\beta=
1.880,1.877,1.874,1871$ (triangles, diagonal crosses, squares and
stars respectively).
Also  shown is the Ising flow (diamonds, bars and fitted curve).
The  rescaling of the gauge block sizes with respect to the standard Ising
scale is $\lambda=0.65$.
 }
 }
\end{center}
\end{figure}

We found  that, among the different gauge couplings used for
$N_T=2$, the supposed critical coupling $\beta=1.880$ is actually ruled out,
no matter which value of the rescaling parameter $\lambda$ is chosen.
\begin{table}
\small
\begin{center}
\begin{tabular}{rllll}
\hline
\hline
&  &  &   &  \\[-3mm]
 \mc{1}{c}{$\alpha$} & \mc{1}{c}{$L_B=3$} &
            \mc{1}{c}{$L_B=4$} &
            \mc{1}{c}{$L_B=5$} &
            \mc{1}{c}{$L_B=6$} \\[0.5mm]
\hline
&  &  &   &  \\[-3mm]
001&\m0.15962(94)~[50]&\m0.17442(52)~[34]&\m0.18333(70)~[27]&
 \m0.18800(92)~[48]\\
011&\m0.02918(57)~[118]&\m0.02800(32)~[4]&\m0.02689(42)~[25]&
 \m0.02500(57)~[58]\\
111&\m0.00668(73)~[103]&\m0.00432(35)~[24]&\m0.00357(45)~[68]&
 \m0.00300(61)~[80]\\
002&$-$0.00728(94)~[177]&$-$0.01282(41)~[62]&$-$0.01514(49)~[77]
 &$-$0.01691(69)~[139]\\
012&$-$0.00277(46)~[70]&$-$0.00444(21)~[12]&$-$0.00500(25)~[56]
 &$-$0.00505(38)~[35]\\
112&$-$0.00115(42)~[62]&$-$0.00177(21)~[5]&$-$0.00179(26)~[59]
 &$-$0.00164(35)~[53]\\
022&\m0.00042(63)~[131]&$-$0.00041(29)~[59]&$-$0.00041(39)~[14]
 &$-$0.00077(49)~[77]\\
122&$-$0.00030(52)~[33]&\m0.00012(21)~[42]&$-$0.00043(30)~[55]
 &\m0.00046(36)~[110]\\
${\phantom{1}9}$&\m0.00085(34)~[31]&\m0.00146(17)~[20]&\m0.00182(21)~[48]&
 \m0.00212(28)~[45]\\
${10}$&\m0.00005(20)~[13]&\m0.00014(10)~[20]&\m0.00014(12)~[11]&
 \m0.00020(16)~[3]\\
${11}$&\m0.00026(55)~[32]&$-$0.00096(28)~[70]&\m0.00006(33)~[23]
 &$-$0.00040(43)~[8]\\
${12}$&$-$0.00013(18)~[6]&$-$0.00013(10)~[9]&$-$0.00013(12)~[11]
 &$-$0.00030(16)~[10]\\
${13}$&$-$0.00020(36)~[9]&$-$0.00020(20)~[15]&$-$0.00029(23)~[41]
 &\m0.00004(31)~[66]\\
${14}$&\m0.00012(59)~[4]&\m0.00012(33)~[5]&\m0.00012(39)~[11]&
 \m0.00033(53)~[95]\\[0.5mm]
 \hline
&  &  &   &  \\[-3mm]
stat &\m\m\m$12\cdot 10^3$& \m\m\m$27\cdot 10^3$&\m\m\m$25\cdot 10^3$&
\m\m\m$20\cdot 10^3$\\[1mm]
\hline
\hline
 \end{tabular}
\parbox[t]{.85\textwidth}
 {
 \caption[Table 4]
 {\label{tab4}
\small
Values of the effective couplings for the $4d$ $SU(2)$ model
at $N_T=2$, $\beta=1.877$ for different block size $L_B$.
In the bottom row the statistics of the last IMCRG iteration
(fixed $L_B$) is given.
Statistical errors are given in parenthesis.
Square brackets contain the change of the coupling
$\Delta K'_\alpha$ in the last IMCRG iteration.
}
}
\end{center}
\end{table}
This can be clearly seen in Figure~\ref{allflow}
where the flows of the NN coupling in
the effective action are shown for the four different gauge couplings
$\beta=1.880,1.877,1.874,1871$ and compared with the fitted curve of the
Ising model.

For $\beta= 1.880$ the system is definitely in the deconfined phase,
whereas the flow for $\beta= 1.871$ moves away  towards the high
temperature fixed point (confinement phase).
The effective coupling values for the best matching trajectory of
$\beta=1.877$ are given in Table~\ref{tab4}.
The less significant values, at $L_B=2$, have been omitted.
We explicitly reported  the statistical errors and the $\Delta
K'_\alpha$ variations in the last IMCRG iteration (the latter in square
brackets).
\par\noindent
Let us conclude this analysis with two remarks.

First, it is worthwhile to stress that weak finite size effects are
present  within this approach: Comparing the $6^3$ block lattice of
the gauge model  with the $8^3$ of the Ising model should not give
sizable systematic errors within  our precision.

As a check, in Table~\ref{tab5} the effective coupling  values of the
ordinary Ising model are reported for two different block lattice sizes,
$L'=8$ and $L'=6$.
The result confirms that all couplings are consistent within errors.
\begin{table}
\small
\begin{center}
\begin{tabular}{lll}
\hline
\hline
  &   &  \\[-3mm]
 \mc{1}{c}{$\alpha$} & \mc{1}{c}{$L'=8$} & \mc{1}{c}{$L'=6$} \\[0.5mm]
\hline
  &   &  \\[-3mm]
 $001$&     \m0.19532(20)&  \m0.19529(60) \\
 $011$&     \m0.02307(6) &  \m0.02315(13) \\
 $111$&     \m0.00172(14)&  \m0.00192(51) \\
 $002$&    $-$0.01899(7) & $-$0.01946(16) \\
 $012$&    $-$0.00551(5) & $-$0.00558(11) \\
 $112$&    $-$0.00168(3) & $-$0.00168(11) \\
 $022$&    $-$0.00002(19)&  \m0.00002(13) \\
 $122$&     \m0.00010(3) &  \m0.00024(29) \\
 $K'_{9}$&  \m0.00195(2) &  \m0.00191(9) \\
 $K'_{10}$& \m0.00024(2) &  \m0.00024(4) \\
 $K'_{11}$&$-$0.00050(3) & $-$0.00049(11)\\
 $K'_{12}$&$-$0.00010(1) & $-$0.00007(9) \\
 $K'_{13}$&$-$0.00014(8) & $-$0.00010(8) \\
 $K'_{14}$&$-$0.00003(7) & $-$0.00025(29)\\[0.5mm]
 \hline
  &   &  \\[-3mm]
stat       &\m\m $1.3\cdot 10^6$& \m\m $3\cdot 10^5$\\[1mm]
\hline
\hline
 \end{tabular}
 \end{center}
\begin{center}
\parbox[t]{.85\textwidth}
 {
 \caption[Table 5]
 {\label{tab5}
\small
Comparison of effective critical couplings for different
sizes of the coarse lattice.
The example shown is the ordinary Ising, with $L_B=9$.
In the bottom line the statistics is given.
}
}
\end{center}
\end{table}

Finally, let us notice that
within our statistic the $\beta=1.874$ flow can also be made compatible
with the Ising trajectory: A better resolution to discriminate between the
two beta values would have required to extend the MC analysis to bigger
block sizes $L_B$, of course with much more CPU time consuming.
Even though, using the block sizes at our disposal the corresponding
fit is not as good as that of the $\beta=1.877$ value.

Therefore we assume the latter as the critical coupling value for
$N_T=2$, consistently (within one standard deviation) with Ref.~\cite{fhk}.
\section{Conclusion and Outlook}

The discussion of MC results shows that
the Svetitsky--Yaffe conjecture is confirmed in a very  fundamental way
by observing matching of the $SU(2)$ RG trajectory with that
of the Ising model.

At the same time, we showed that  IMCRG works well as a method to compute the
effective action of Ising type degrees of freedom in a genuine
non--Ising model like $4d$ finite temperature $SU(2)$ gauge theory.

Notice also
that this kind of calculations could be done on workstations, with relatively
small computer resources.

An extension to $N_T$ greater than two  would be interesting but  more
expensive.  The reason is that with increasing temporal size  the small
$L_B$ actions move farther away from the fixed points, i.e.\ they need
to be blocked more in order to come close to the reference  Ising
flows.  This observation is in agreement with the fact that also in
more standard approaches, e.g.\ via the Binder cumulant,  the spatial
size of the lattice has to be increased very much with increasing $N_T$.

Finally, it would be of interest to check this approach with different
blocking prescriptions.
The rate of approaching the RG fixed point is in fact very sensitive to
the blocking rule used and a faster convergence can in principle be
obtained using a more sophisticated blocking scheme than the majority rule.

\section*{Acknowledgements}
We would like to thank M.~Caselle for interesting exchange of opinions.
Access to CPU resources of the Torino I.N.F.N. is gratefully acknowledged.

\end{document}